\documentclass[prd,preprint,nofootinbib,tightenlines]{revtex4}
\usepackage{amsmath}
\def\bea{\begin{eqnarray}}
\def\eea{\end{eqnarray}}
\def\beq{\begin{equation}}
\def\eeq{\end{equation}}
\def\Eq#1{Eq.\ (\ref{#1})}
\def\Re{\;{\rm Re}\,}
\def\Im{\;{\rm Im}\,}
\def\A{{\cal A}}

\begin{document}

\title{Decoherence due to the Horizon after Inflation}
\author
   {%
    Jonathan W.\ Sharman and Guy D.\ Moore
   }%
\affiliation
   {%
    Department of Physics,
    McGill University, 3600 rue University,
    Montr\'eal QC H3A 2T8, Canada
    }%
\date{July 2007}

\begin{abstract}
The fluctuations in the inflaton field at the end of inflation which
seed the density perturbations are prepared in a pure quantum state.  It is
generally assumed that some physics causes this pure state to decohere
so that it should be treated probabilistically.  We show that the
entanglement entropy between the universe inside our observable horizon
and that outside our horizon is sufficient to do this.  For
the modes which are super-Hubble at the end of inflation, this
entanglement entropy grows with volume inside the horizon, rather than
with the horizon's area, and is proportional to the number of e-folds
since Hubble crossing.
\end{abstract}

\maketitle

\section{Introduction}

Inflation provides an elegant description for the large scale appearance
of our universe (very large, approximately homogeneous and isothermic,
and spatially flat to within measurement errors)
\cite{Guth,Linde,Albrecht}.  It also provides a mechanism to explain how
the density perturbations which seeded structure formation (and
therefore us) came to be (for a comprehensive review see
\cite{BrandenbergerMukhanov}).  This occurs because, during inflation,
the quantum state of (the approximately harmonic oscillator $k$-modes
of) the inflaton field become highly squeezed \cite{joyce}.

This squeezing, present on super-Hubble
scales at the end of inflation,
means that the inflaton is effectively taking on different values at
widely separated points in space, which leads to an inhomogeneity in the
temperature after reheating (or certain nonzero metric elements in a
different gauge choice).

One puzzle in this picture is that the state of the inflaton field is a
quantum state.  It is not really correct to say that it is ``ahead'' at
some points and ``behind'' in others; its quantum state
contains amplitudes to be ahead and behind at different points.  Usually
in treating the subsequent evolution of the system one replaces this
quantum state with a classical probability distribution or ensemble with
the same statistical properties.  This artificially introduces a large
entropy into the state, since a pure quantum state is being replaced
with a statistical ensemble.  It is an interesting question whether this
procedure is really legitimate, and if so how it is justified.

This issue has been mostly resolved by Polarski {\it et al}, who showed
\cite{polarski} that the difference between performing calculations
using the quantum squeezed states and the classical probabilistic
distribution is exponentially suppressed in the number of e-folds of
inflation.  However it is still an interesting question whether the
statistical treatment is in some way better justified than this.  Are we
actually {\em compelled} to consider the primordial fluctuations
probabilistically?

Burgess {\it et al} have argued that decoherence of the quantum state
actually occurs due to interactions between the long-wavelength modes of
the inflaton fields and shorter wavelength modes \cite{cliff}.  Assuming
that these shorter wavelength modes are not observed, the description of
the long wavelength modes should be made by integrating them out,
resulting in an entanglement entropy.  The calculation depends on the
details of the interactions between long and short wavelength modes and
on the non-observation of the short wavelength modes.

We believe that it is unnecessary to consider this division of modes
into long and short wavelength to show that the universe we observe will
possess a large entanglement entropy and must be described
statistically.  This is because the post-inflationary universe has a
horizon.  We can only make observations of our universe within
our horizon patch.  But inflation sets up correlations between our
causal patch and the universe ``outside'' that patch.  Indeed, the basis
which diagonalizes the wave functional at the end of inflation is the
momentum basis, which is maximally nonlocal.
Since it is impossible to observe the
universe outside our horizon, we are obliged to integrate it out.  The
goal of this paper is to investigate the degree of entanglement with the
universe outside our horizon.  In particular we will compute the entropy
of entanglement.

Our method is based on that used by Srednicki \cite{entropyandarea}, who
considered the entanglement entropy associated with cutting a spherical
region out of space, as a toy model of a black hole's entropy.  We
extend his treatment to highly squeezed quantum states.  Our main
conclusion is that, for the highly squeezed case, the entanglement
entropy grows with the volume of the visible universe, rather than with the
surface of the boundary between the observed and unobserved regions, as
occurs for vacuum fluctuations.
The entropy per mode scales with the log of the conformal time,
which is the same as the number of e-foldings of inflation.  This
saturates a bound placed by Kiefer {\it et al} \cite{kiefer} on the
entropy per mode which is allowed without destroying the phase
information in the fluctuations, responsible for the acoustic peaks.
Technically this conclusion applies only for the very longest
wavelengths; if inflation took $n$ e-folds to stretch our observed
universe out of one Hubble patch, then the volume behavior we find is
valid for fluctuations with wavelength more than $1/n$ of the current
horizon length.
(This volume dependence is consistent with the entropy conjecture of Bousso
\cite{Bousso} because it applies only to the few infrared modes which
were super-Hubble at the end of inflation and so became highly
squeezed. The horizon area
behavior argued by Bousso is dominated by the much more numerous shorter
wavelength modes.)

The entropy per mode which we compute is an inevitable result of the
horizon structure of the post-inflationary universe and does not rely on
any assumptions about the interactions of the inflaton field with other
(short-wavelength) degrees of freedom.  It is therefore robust, and
sufficient to explain the statistical nature of the universe we observe.

\section{Squeezed States}

A squeezed harmonic oscillator state is generated from the vacuum state
by a squeezing operator, which is a function of two variables, the
amplitude of squeezing $r$ and the phase $\Phi$.  The squeezing operator
for a momentum-space mode of wave vector $k$ of a quantum field is
generically by \Eq{sqeezing}, below.

During inflation, the state of a noninteracting, light scalar field can
be described by the quantum state of each momentum-space state.  The
cosmological expansion causes the momentum space states with wavelengths
longer than the Hubble scale become highly squeezed, in a manner which
has been worked out by Albrecht {\it et al} \cite{joyce}.
They find that the evolution equations for $r_k$ and $\Phi_k$ in terms
of proper time $\eta$ are

\begin{eqnarray}
U & = & S(r_k,\theta_k) R(\Theta_k),
  \\
S(r_k,\Phi_k) & = & \exp \left[ \frac{r_k}{2}
    \left( a_k a_{-k} e^{-2i \Phi}-a_k^\dagger a_{-k}^\dagger e^{2i \Phi}
    \right)  \right] \label{sqeezing},
  \\
R(\Theta_k) & = & \exp \Big[ -i\Theta_k(a_k^\dagger a_k
                     + a_{-k}^\dagger a_{-k}) \Big],
  \\
r_k & = & -\sinh^{-1}\left( \frac{1}{2k |\eta|} \right) \label{rofk},
  \\\label{contestedsigh}
\Phi_k & = & \left( -\frac{\pi}{4} +\frac12
          \arctan{ \frac{1}{2k |\eta|}}\right) \label{phiofk},
  \\
\Theta_k & = & -k|\eta| -\arctan{ \frac{1}{2k |\eta|}} \,.
\end{eqnarray}
Here $U$ is the time evolution operator of the inflaton and
$R(\Theta_k)$ is an overall phase.
Note that these solutions are for the exponentially expanding deSitter
stage, in the Bunch Davies vacuum. 

Evaluating the state of this $k$-mode in the field basis gives
\cite{wavejustification}
\bea
\Psi [\phi(k,\eta),\phi(-k,\eta)] 
    &=&\langle \phi(k,\eta), \phi(-k,\eta) | S | 0 \rangle \nonumber \\  
& = & N_k   \exp(  -\lvert \phi(k,\eta) \rvert^2 \Omega_k   )
  \\ \label{omegak}
\Omega_k & = & k \frac{1+i\sin(2\Phi_k)\sinh(2 r_k) }
   {\cosh(2 r_k)-\cos(2\Phi_k)\sinh(2r_k) } \nonumber\\
 &=&k\left(\frac{k^2 |\eta^2| +ik|\eta|}{1+k^2 |\eta|^2} \right)\,.
\eea
Here $\phi(k,\eta)$ is the scalar field,  called $y(k,\eta)$ in
\cite{wavejustification}.  $N_k$ is a
normalizing constant; we will not keep track of such normalizations,
ensuring the correct normalization at the end by scaling the density
matrix to have trace 1. 
The fact that the wavefunction can be put in this simple, Gaussian form
means that we can adapt the approach Ref.\ \cite{entropyandarea} used to
calculate the entropy of a sphere of space to calculating the entropy
of the squeezed states. 

\section{Integrating out position-space modes}

As discussed, we want to compute the entropy of entanglement of the
visible part of the universe (that which lies within our horizon) when
the invisible part (space outside our horizon) is integrated out.  We are
only interested in the behavior and entropy of relatively long
wavelength modes, those responsible for the density perturbations on
scales which are linear today or can be probed at a time scale when they
were linear.  At the end of inflation and in comoving coordinates, these
modes had wavelengths much larger than the Hubble scale and so were all
highly squeezed.  To describe only these long wavelength modes,
we should implement some infrared regulator,
which for convenience we take to be a lattice discretization of space.
(We do not think that the choice of regulator is important to our
results.)  We will also make our system finite by imposing boundaries
far outside the observable horizon.  We will seek the limit where the
boundaries are moved to infinity and will study the dependence on the
infrared regulation.  As a further simplification we consider a
3-dimensional system where the visible universe
is a cube rather than a sphere; again the qualitative behavior should be
the same as for the spherical case.  As a test case we also consider a
1-dimensional system.

This effectively reduces the problem to a quantum mechanical problem of
the product of a large number of harmonic oscillators.  We will find the
wave function and from it the density matrix, which is most easily done
in the momentum basis.  Then we will transform to the position basis and
integrate over the coordinates outside the horizon.  The procedure is
similar to that of Srednicki \cite{entropyandarea}, except that we
consider highly squeezed SHO states, rather than vacuum states.

In one dimension, the discretized box Hamiltonian for a free scalar
field is
\bea
H&=&\sum_{j=1}^{N}  \left[ -\frac{1}{l^2} \phi(x_j)
  [\phi(x_{j+1})-2\phi(x_j)+\phi(x_{j-1})]+\pi_j^2 \right] \\ \nonumber
&=&\frac12 \pi^\dagger \pi+\frac12 \phi(x)^\dagger M \phi(x),
\eea
where $l$ is a lattice spacing, $Nl$ is the extent of the universe, and
$\phi(x_j)$ and $\pi_j$ are the $N$ generalized coordinates and conjugate
momenta.  Here $(\phi_{j+1}-2\phi_j+\phi_{j-1})$ is a discrete
approximation of $l^2 \nabla^2 \phi$, and we switch to matrix notation
in the last expression.  We take $\phi(0)=0=\phi(N+1)$, which enforces
Dirichlet boundary conditions.  This choice of boundary conditions
should be immaterial if we take an appropriate large volume limit.  We
have tested this by using mixed Dirichlet-Neuman and antiperiodic
boundary conditions.

The matrix $M$ is diagonalized by a sine transform.  Explicitly, 
defining the matrix $U$ as
\beq
U_{jk} = \sqrt{\frac{2}{N+1}}\; \sin \left(\frac{j k\pi}{N+1} \right)\, , 
\eeq 
the matrix $M$ becomes $M = U^\dagger \omega^2 U$, with
\beq
\omega = {\rm Diag}[\omega_j] \, , \qquad
\omega_j^2 = \frac{4}{l^2} \sin^2\left(\frac{\pi j}{2(N+1)}\right)\,.
\eeq

The wave function for the scalar field is easiest to express in this
momentum basis.  Defining $\phi_j(k) = U^\dagger_{jl} \phi_l(x)$, then
up to a normalization constant, the squeezed state in the $k$ space field
basis is
\bea
\langle \phi | S \rangle &=& \prod_{k=1}^N \langle \phi_k | S_k
\rangle\\ \nonumber
&=&  \prod_{k=1}^N \exp(-\Omega_k \phi_k^2)\\ \nonumber
&=& \exp(-\phi(k)^\dagger \Omega \phi(k)),
\eea
where $\phi(k)$ is a vector of $N$ $k$ modes and $\Omega$ is the
diagonal matrix with 
elements $\Omega_k$ given by \Eq{omegak} with $\omega_k$ replacing $k$. 

We can express this in the position basis using the unitary
transformation matrix $U$ found earlier.  Defining
${\cal N} = U^\dagger \Omega U$, the state in position basis is
\beq
\langle \phi | S \rangle = \exp\,-\phi(x)^\dagger {\cal N} \phi(x) \,,
\eeq
giving the density matrix
\bea
\rho(\phi(x),\phi(x)') & = &
   \langle \phi | S \rangle \langle S|\phi' \rangle \nonumber \\
 & = & \exp- \left(\phi(x)^\dagger {\cal N} \phi(x)
   +\phi(x)^{' \dagger}{\cal N}^\dagger \phi(x)' \right) \,.
\eea

With the density matrix in this form it is now possible to calculate the
partial trace of the physical modes outside the visible universe using
an approach similar to that of \cite{entropyandarea}.  However before
continuing it is important to note that the method used there must be
altered slightly because ${\cal N}$ is not Hermitian, that is, $\Omega$
is not real.

Now we subdivide the $N$ lattice points into $m$ points making up the
visible universe and $n=N-m$ modes lying outside our horizon, which we
therefore want to integrate
out. In the 1-D case the $m$ modes are taken to be bounded on each side
by $\frac{n}{2}$ modes, and in the 3-D case the visible universe is
taken to be the central modes of an $N\times N \times N$ cube.

We should therefore write the matrices $U$ and $\Omega$ in block form as
follows: 
\bea \label{umatrix}
U&=&\left(\begin{array}{cc}
U_1&U_3\\
U_4&U_2
\end{array}
\right),
\\
\label{betamatrix}
\Omega&=&\left(\begin{array}{cc}
\Omega_n&0\\
0&\Omega_m
\end{array}
\right).
\eea
From here it is straightforward to see that
\beq
\phi^\dagger {\cal N} \phi = \begin{array}{cc}
\big[ y^\dagger&x^\dagger \big] \\ &
\end{array} \!\!
\left[\begin{array}{cc}
A_R+iA_c&C_R+iC_c\\
C_R^\dagger+iC_c^\dagger&B_R+iB_c
\end{array} \right]
 \left[\begin{array}{c}
y\\
x
\end{array} \right],
\eeq
Where
\bea
A_R&=&U_1^\dagger \Re[\Omega_n] U_1+U_4^\dagger  \Re[\Omega_m] U_4 \label{ar}\\
A_c&=&U_1^\dagger \Im[\Omega_n] U_1+U_4^\dagger  \Im[\Omega_m] U_4 \label{ac}\\
C_R&=&U_1^\dagger \Re[\Omega_n] U_4+U_4 ^\dagger \Re[\Omega_m] U_2 \\
C_c&=&U_1^\dagger \Im[\Omega_n] U_4+U_4 ^\dagger \Im[\Omega_m] U_2 \\
B_R&=&U_3^\dagger \Re[\Omega_n] U_3+U_2^\dagger  \Re[\Omega_m] U_2 \label{br}\\
B_c&=&U_3^\dagger \Im[\Omega_n] U_3+U_2^\dagger  \Im[\Omega_m] U_2 \label{bc}\\
B&=&B_R+iB_c.
\eea

Note that $A_{R,c},B_{R,c}$ are all Hermitian matrices and we have split the
$N$ modes into a $y$ vector of $n$ modes and an $x$ vector of $m$
modes.  We will be tracing over the $y$ modes.  The partial trace of a
density matrix can be found by integrating over the appropriate modes;
that is,
\beq
\rho_{\rm proj}(x;x') = \int_{-\infty}^\infty dy \; \rho(x,y;x',y).
\eeq
Replacing $y'$ with $y$, the density matrix to be integrated is
\bea
\rho = \exp- && \!\!\!\!
  \Big[2y^\dagger A_R y +y^\dagger [C_R(x{+}x')+iC_c(x{-}x')] 
 +[(x{+}x')^\dagger C_R^\dagger +i(x{-}x')^\dagger  
  C_c^\dagger ]y  
  \nonumber   \\ &&
  +x^\dagger B x +x'^\dagger B^\dagger x'  \Big] \,.
\eea
The Hermitian matrix $A_R$ can be written as a unitary transformation of
a diagonal matrix, $A_R \equiv V^\dagger \lambda V$.  After
completing the square for $y$ and carrying out the Gaussian $y$ integration,
the density matrix becomes
\bea
\rho_{\rm proj}&=&\exp-\Big( x^\dagger (K-J-i(G+G^\dagger) +B) x
  +x^{' \dagger} (K-J+i(G+G^\dagger) +B^\dagger)x' \nonumber \\ \nonumber
&&-x^\dagger(K+J+i(G-G^\dagger))x' -x^{' \dagger}(K+J-i(G-G^\dagger))x
\Big)  \\
&\equiv& \exp- \left( x^\dagger \Gamma x+x^{' \dagger} \Gamma^\dagger x' 
  -x^\dagger \delta_1 x' -x^{' \dagger}\delta_2 x   \right),
\eea
where
\bea
G&\equiv&C_c^\dagger V^\dagger \frac{\lambda^{-1}}{2} VC_R \\
K&\equiv&\left[VC_c+(VC_c)^{*}\right]^\dagger 
      \frac{\lambda^{-1}}{8}\left[VC_c+(VC_c)^{*} \right]  \\
J&\equiv& C_R^\dagger V^\dagger \frac{\lambda^{-1}}{2} VC_R \\
\Gamma&\equiv& K-J-i(G+G^\dagger) +B \\
\delta_1&\equiv&K+J+i(G-G^\dagger) \\
\delta_2&\equiv&K+J-i(G-G^\dagger) \, .
\eea
In the case where $\Omega$ is purely real, this reduces to the result
of \cite{entropyandarea}. 

\section{Entropy, eigenvalues, and eigenvectors}

\subsection{Diagonalization of $\rho$ into normal modes}

We now have the density matrix for the visible part of the universe.
The question is, can this be represented by a stochastic distribution,
or is its quantum nature still observable?  To determine this we
calculate the amount of decoherence of the density matrix by computing
its entropy.  To do that we need to find the eigenvalues and
if possible the eigenvectors of the density matrix.

Since $\delta_1$ and $\delta_2$ are both Hermitian, we can further
simplify the mixed term in the density matrix:
\bea
x^\dagger \delta_1 x' +x^{' \dagger}\delta_2 x&=&z^\dagger
\lambda_{\delta 1} z'+z^{' \dagger}Q \delta_2Q^\dagger z\\ \nonumber
=z^{' \dagger} \lambda_{\delta 1} z+z^{' \dagger}Q \delta_2Q^\dagger
z&=&x^{' \dagger} (\delta_1+\delta_2)x\\ \nonumber
&=&2x^{' \dagger} (K+J)x,
\eea
where $Q^\dagger \delta_1Q= \lambda_{\delta 1}=$diagonal, and $x=Qz$.

Now we split $\Gamma$ into its Hermitian and anti-Hermitian parts:
\bea
\Gamma&=&\A+iD,\\
\A&=&K-J+B_R, \label{realgamma} \\
D&=&-G-G^\dagger +B_c.
\eea
This gives us a density matrix of the form:
\beq
\rho_{\rm proj} = \exp - \left[( x^\dagger \A x+x^{' \dagger} \A x')
  +i( x^\dagger D x-x^{' \dagger} Dx')
 -2x^\dagger(\A-(B_r-2J))x'   \right]  .
\eeq
We know that $\A$ is positive definite since the density matrix is
normalizable.  Diagonalizing $\A=W\Lambda W^\dagger$ and
dividing by $\sqrt{\Lambda}$ gives
\beq
\rho_{\rm proj} = \exp - \left[( u^\dagger u+u^{' \dagger} u')+i(u^\dagger D'
  u-u^{' \dagger} D'u')
-2u^\dagger(1-\Lambda^{-\frac12}W^\dagger(B_R-2J)W\Lambda^{-\frac12})u'
 \right] \,,
\eeq
 where $x=\Lambda^{1/2}Wu$.
Diagonalizing 
 $\Lambda^{-\frac12}W^\dagger(B_R-2J)W \Lambda^{-\frac12}$ gives
\bea
\rho_{\rm proj}&=&\left( \prod_{j=1}^m \sqrt{\frac{\epsilon_j}{\pi}}  \exp(-
    [( v_j ^2 +v_j^{'2})-2v_j(1-\epsilon_j)v_j'   ] )  \right) \exp(-i(
    v^\dagger D'' v-v^{' \dagger} D''v')) \nonumber \\
&\equiv& \left( \prod_{j=1}^m \rho_j (v_j;v_j' )  \right) \exp(-i( v^\dagger
    D'' v-v^{' \dagger} D''v')) \, ,
\label{final_rho}
\eea
where $v$ is in the new diagonal basis, and $\epsilon_j$ are the
eigenvalues of
$\Lambda^{-\frac12}W^\dagger(B_R-2J)W \Lambda^{-\frac12}$.  
The second equation defines
$\rho_j (v_j;v_j' )\equiv \sqrt{\frac{\epsilon_j}{\pi}}  \exp(- [( v_j
^2 +v_j^{'2})-2v_j(1-\epsilon_j)v_j'   ] )$.

\subsection{Proof of unimportance of complex phase}

Besides the $D''$ term, this density matrix in \Eq{final_rho} is in the
same form as found in \cite{entropyandarea}.  We now show that the $D''$
term does not change the eigenvalues and only modifies the eigenvectors
in a trivial way.

Write the eigenfunctions of $\rho(v_i,v_i')_i$ as $f_{n(i);i}$,
satisfying
\beq
\int_{- \infty}^{\infty} \rho_i (v_i,v_i')f_{n(i);i}(v_i')dv_i' =
p_{n(i);i} f_{n(i);i}(v_i) \,,
\eeq
with $p_{n(i);i}$ the associated eigenvalue.
Then the eigenvectors and eigenvalues of our density matrix are a $D$
dependent phase factor times products of these, namely
\bea
g_{n(1),n(2),..,n(m)}(v)
  & = & e^{(-iv^\dagger D'' v)} \prod_{j=1}^m f_{n(j),j}(v_j) \, , \\
p_{n(1),n(2),..,n(m)}&=&\prod_{j=1}^m p_{n(j);j} \, .
\eea
This is easy to check:
\bea
\int_{-\infty}^\infty \!\! dv' \rho_{\rm proj}(v,v') g(v')
  & = & \int_{-\infty}^\infty \!\! dv' 
    \left( \prod_{j=1}^m \rho_j (v_j;v_j' ) \right) 
    e^{(-i( v^\dagger  D'' v-v^{' \dagger} D''v'))} 
    e^{(-iv^{'\dagger} D'' v')} 
    \prod_{j=1}^m f_{n(j),j}(v_{j}') \nonumber \\
 & = & \exp(-i v^\dagger D'' v)
    \int_{-\infty}^\infty dv' \left( \prod_{j=1}^m \rho_j (v_j;v_j' ) 
      f_{n(j),j}(v_{j}')  \right) \nonumber \\
 & = & \exp(-i v^\dagger D'' v) 
    \left( \prod_{j=1}^m p_{n(j)}f_{n(j),j}(v_{j})  \right) \nonumber \\
 & = & p_{n(1),n(2),..,n(m)}g(v) \,.
\eea
Therefore the complex phase does not actually affect the eigenvalues of
the density matrix, and only adds an overall phase to the eigenvectors.
However this complex phase will be
important when taking further partial traces. 

\subsection{Entropy per mode}

It was shown in \cite{entropyandarea} that the density matrix
$\rho_j(v_j;v_j')$ found in \Eq{final_rho} has eigenvectors and
eigenvalues of
\bea
f_{n}(v_j) & = & \frac{1}{\sqrt{\alpha n! 2^n
    \sqrt{\pi}}}H_{n}\left( \alpha^{-\frac12} v_j \right)
     \exp \left( - \frac{\alpha}{2} v_j^2 \right) \, , \\
p_{n} & = & (1-\xi)\xi^n \,, \\
\alpha & = & \sqrt{2\epsilon-\epsilon^2}\,,\\
\xi & = & \frac{1-\epsilon}{1+\alpha} \,.
\eea
In other words, up to the phase factor which we just showed is
irrelevant, there exists a basis in which the density matrix behaves
like the product of density matrices for individual harmonic
oscillators, with Poisson distributed occupancies of states which are
the same as the conventional harmonic oscillator number states.

The entropy is the
sum of the entropies for the normal mode density matrices.
The entropy of the Poisson distribution described by $p_n =
(1-\xi)\xi^n$ is
\beq
s_\xi = -\log(1{-}\xi) - \frac{\xi}{1{-}\xi} \log(\xi) \,.
\eeq
As we will see, we will be interested in the limit where $\epsilon_j \ll
1$, in which case $\xi \simeq 1-\sqrt{2\epsilon}$ and the entropy is
\beq
S = \sum_j s_{\xi_j} \simeq 
   \sum_j \left( 1 - \frac{1}{2} \log (2\epsilon_j) \right) \,.
\eeq

\subsection{Large Squeezing Limit}

For large squeezing, $\Omega_k \simeq k(k^2 \eta^2+ik|\eta| )$, which is
small and dominated by the imaginary part.
The overall scale of $\Omega_k$ is immaterial;
multiplying all $\Omega_k$ by a common factor scales out in determining
the $\epsilon_j$ and therefore the entropy.  However, the $k$ dependence
and the relative size of the
real and imaginary parts is relevant. Since $k\eta$ is exponentially
small in the number of e-foldings, we should study the
scaling of the eigenvalues $\epsilon_j$ in the small $k|\eta|$ limit.

Ultimately the entropy depends on the sizes of the eigenvalues
$\epsilon_j$ of the matrix 
$\Lambda^{-\frac12}W^\dagger(B_R-2J)W \Lambda^{-\frac12}$
in \Eq{final_rho}.  Therefore we should study the scaling behavior in
$\eta$ of the pieces of this matrix.
It is convenient to multiply $\Omega_k$ by $l/\eta$, with $l$ the lattice
spacing (shortest wavelength under consideration), so
\bea\label{omegaofeta}
\Omega_k(\eta) \rightarrow 
   k\left( (l k)^2 \left( \frac{|\eta|}{l} \right)+i(l k) \right).
\eea
In the small $\frac{|\eta|}{l}$ limit, provided that the eigenvalues of $K$ are
all nonzero, one may neglect $B_R-J$ in \Eq{realgamma}.  In this case
the $\eta$ dependence scales simply; the matrix
$\Lambda^{-\frac12}W^\dagger(B_R-2J)W \Lambda^{-\frac12}$, and therefore
the $\epsilon_j$, scale as $\left| \frac{\eta}{l} \right|^2$.  Defining
$\epsilon(1)_j$ as the set of eigenvalues arising by setting
$\frac{|\eta|}{l}=1$ and setting $\A=K$ in \Eq{realgamma}, the approximate
values of the $\epsilon_j$ are
\beq
\epsilon_j = \frac{\eta^2}{l^2} \epsilon(1)_j \,.
\eeq
Therefore the entropy is approximately
\beq
S \simeq \sum_j \ln \frac{l}{|\eta|} + 1-\frac12 \ln2 
  -\frac12 \ln \epsilon(1)_j \,.
\label{eq:s}
\eeq
This demonstrates that, for $\eta/l$ parametrically small (modes which
were well outside the horizon at the end of inflation), the entropy is
extensive in the number of modes and scales as the number of e-folds by
which the mode was outside the horizon (the log of the level of
squeezing).

Note however that this conclusion relies on the $\epsilon(1)_j$ not
being exponentially large.  Some of these eigenvalues could potentially
be exponentially large if there is
another parametrically large number in the game, such as the number $m$
of points inside the horizon.  Therefore our conclusion that the entropy
scales as the volume and as $-\ln |k\eta|$ is only robust if we consider
fluctuations on very long scales, $m < -\ln |\eta/l|$.

We have been unable to find a closed-form solution for the eigenvalues
$\epsilon(1)_j$.  Instead we have performed numerical experiments to check
how the residual entropy in \Eq{eq:s} scales with volume.
Recall that the choice of the number of points inside the
visible universe corresponds to a selection of what scales of
perturbations to consider, while we should seek the limit that the
volume outside the visible universe becomes large.

We studied the entropy in both the one dimensional and three dimensional
cases.  The matrix manipulations rapidly become numerically unstable and
were only able to achieve a large value of $m$ for the
1-dimensional case.  For small $m$ (an analysis only considering modes
of near-horizon wavelength), we find as expected that all eigenvalues
$\epsilon(1)_j$ are $O(1)$ and therefore the entropy grows with the
number of modes and scales with the number of e-folds.  When we increase
$m$, we observe that the mean value $\frac{1}{m} \sum_{j=1\ldots m}
\ln \epsilon(1)_j$ grows roughly linearly with $m$; the largest
eigenvector $v_j$ is one which avoids the boundary.  This suggests
(though we have not been able to show it analytically) that the volume
behavior we obtain will break down when one considers wavelengths
shorter than the current horizon by more than the number of e-foldings
of inflation.

Return to the longest wavelength modes, where we have shown that the
entropy per mode is $\simeq n_{\rm efolds}$.
This is exactly what
Kiefer {\it et al} \cite{kiefer} predicted to be the upper bound on the
entropy of the modes.  They argue that if the entropy were any higher,
then the experimentally observed acoustic oscillations of the CMB
fluctuations would not be seen.  Therefore, so far as the modes which
can actually be observed are concerned, the decoherence arising from
entanglement with the (unobservable) region outside our horizon is
sufficient to force a probabilistic description of the density
fluctuations, regardless of whether or not there are additional
mechanisms for decoherence.

\section{conclusion}

The post-inflationary causal structure of the Universe has a horizon,
meaning that we can only observe physics in some finite domain of the
universe.  However, inflation induces a large entanglement between the
IR field modes in this region and those outside our horizon.  We have
calculated the entanglement entropy for the IR modes responsible for
density perturbation seeds.  The entropy scales with the number of modes
in the visible universe, that is, with the volume of the universe times
the cube of the $k$-vector at which density perturbations are measured.
The entropy per mode obeys the simple rule $s\approx r$,
where $r$ is the squeezing factor (the number of e-foldings of
inflation during which the relevant mode was outside the Hubble radius,
of order 60).  This far exceeds the cosmic variance limit on the amount
of information available in the density perturbations.  In fact, it
saturates the upper bound predicted by Kiefer {\it et al} \cite{kiefer}.
This result holds for the longest wavelength modes, those with
$H_0 \lambda > 1/r$, with $H_0$ the current Hubble constant (roughly the
inverse of the horizon length).

This entanglement entropy is large enough to render unmeasurable any
quantum coherence in the initial squeezed state set up by inflation and
to justify a probabilistic description of the initial seeds of the
density fluctuations.

\pagebreak

\noindent{\bf acknowledgements}

\medskip

We would like to thank Cliff Burgess, Nima Lashkari, and Robert
Brandenberger for useful conversations.
This work was supported in part by
the Natural Sciences and Engineering Research Council of Canada.

\end{document}